\documentclass[pre,twocolumn]{revtex4}
\usepackage{graphicx}
\usepackage{amsmath,amssymb}

\begin{document}
\title{Twirling motion of actin filaments in gliding assays with
  non-processive myosin motors} 
\author{Andrej Vilfan}
\affiliation{J. Stefan
    Institute, Jamova 39, 1000 Ljubljana, Slovenia, andrej.vilfan@ijs.si}
\begin{abstract}We present a model study of gliding assays in which actin
  filaments are moved by non-processive myosin motors.  We show that
  even if the power stroke of the motor protein has no lateral
  component, the filaments will rotate around their axis while moving
  over the surface.  Notably, the handedness of this twirling motion
  is opposite from that of the actin filament structure.  It stems
  from the fact that the gliding actin filament has ``target zones''
  where its subunits point towards the surface and are therefore more
  accessible for myosin heads.  Each myosin head has a higher binding
  probability before it reaches the center of the target zone than
  afterwards, which results in a left-handed twirling. We present a
  stochastic simulation and an approximative analytical solution. The
  calculated pitch of the twirling motion depends on the filament
  velocity (ATP concentration).  It reaches about 400nm for low speeds
  and increases with higher speeds.
\end{abstract}
\maketitle

\section*{Introduction}

\begin{figure}
\begin{flushleft}
\raisebox{4cm}{(A)}\hspace*{-2em}\includegraphics*{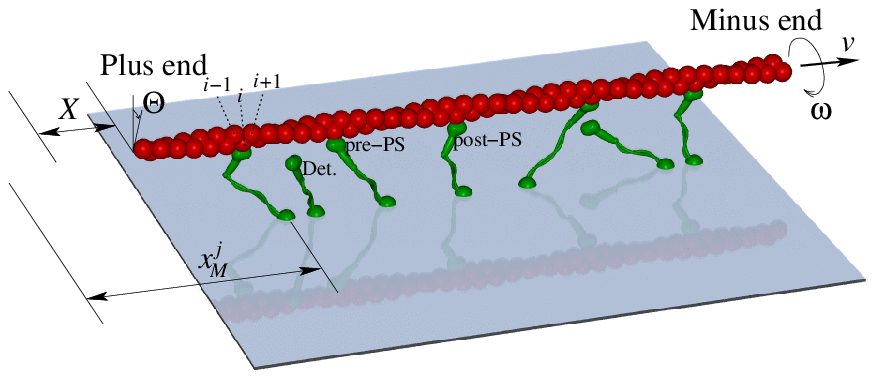}\\
\raisebox{2.2cm}{(B)}\hspace*{-2em}\includegraphics*{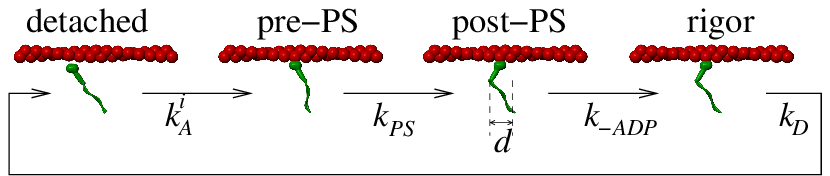}\\
\raisebox{4cm}{(C)}\includegraphics*{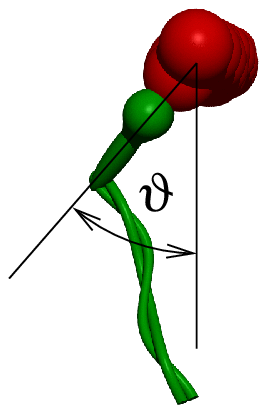}\\
\raisebox{2.5cm}{(D)}\includegraphics*{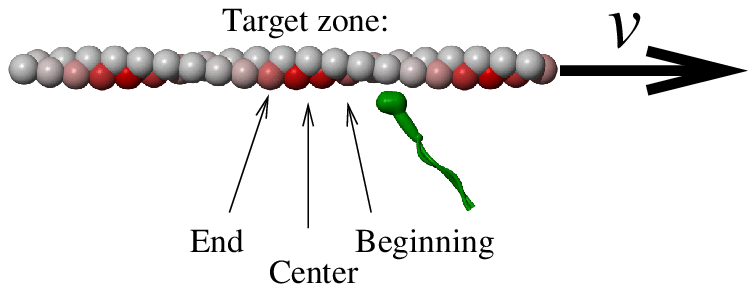}
\end{flushleft}
\caption{(A) Model definition. The linear motion and rotation of the actin
    filament are denoted with $X$ and $\Theta$. Motors are situated underneath
    the actin filament at randomly distributed positions $x_M^j$ and each motor
    can be in the detached, pre-powerstroke or post-powerstroke state. (B) Duty
    cycle of a myosin head. The head binds to the actin site $i$ with rate
    $k_A^i$, then quickly undergoes the power stroke with rate $k_{PS}$,
    releases ADP with rate $k_{-ADP}$ and detaches with rate $k_D$. The step
    size is $d=8\,\rm nm$. (C) Front view of a bound head. For the sake of
    simplicity, we assume that the azimuthal component of the elastic energy
    needed to bind to site $i$ only depends on the angle $\vartheta$. (D)
    Target zones on an actin filament. When a filament is moving past an actin
    head, it is more likely to bind to submits pointing towards the
    myosin-covered surface.  Red color denotes the sites with the highest
    binding rate. Besides that, sites at the beginning of a target zone have a
    higher binding probability than those at its end.}
\label{fig:1}
\end{figure}

Gliding assays, also known as motility assays, represent the oldest in vitro
technique to study motor proteins \cite{toyoshima87,howard89}. They consist of
attaching motors (like myosins or kinesins) with their tails to a glass surface
and adding the filaments (actin or microtubules).  The motors will then pull
the filaments and make them glide over the surface 
(Fig.~\ref{fig:1}A).  Gliding assays are the most convenient way of
testing motors for their functionality, measuring their speed in the absence of
load and for testing their processivity. Several experimental and theoretical
studies were dealing with the pathways of such filaments in the two-dimensional
plane \cite{duke95,bourdieu95,Gibbons.Jose2001}. Interestingly, one group
observed that gliding actin filaments move in a helical fashion
\cite{Tanaka.Ishiwata1992,nishizaka93}.  In a subsequent experiment the pitch
of rotation was determined as about $1\,\rm \mu m$, although the applied
optical detection method did not allow discrimination between left- and
right-handed rotation \cite{Sase.Kinosita1997}.

Helical motion of myosin motors has been very important in a somewhat
different context.  The processive motor myosin V has an average step
size that is close, but not precisely equal to the actin periodicity.
The helical motion of a motor around the actin filament therefore
presents a very accurate way of measuring the difference between its
step size and the filament pitch.  Ali and coworkers have observed
that myosin V walks on an actin filament along a left-handed helix
with a pitch of $2.2\,\rm \mu m$ \cite{Ali.Ishiwata2002} and thus has
a step size slightly shorter than the actin half-pitch (for a
discussion of the myosin V step size see
\cite{Vilfan2005,vilfan2005b}).

Myosin VI, despite having a shorter lever arm than myosin V, showed
either straight walking, or, in $20\%$ of cases, a helical path with a
pitch of $2.3\,\rm \mu m$ \cite{Ali.Ikebe2004}.  Sun et
al. \cite{Sun.Goldman2007} confirmed this result, but also showed that
the relatively straight motion contains a large amount of random
wiggling. New experiments on myosin X also show a left-handed helical
motion with a pitch that is somewhat shorter than that of myosin V and
VI \cite{Arsenault.Goldman2009}.

In a recent experimental study Beausang and coworkers
\cite{Beausang.Goldman2008} used polarized total internal reflection
microscopy to study the twirling motion of actin filaments in gliding assays
with processive myosin V and non-processive muscle myosin (myosin II).  While
the twirling of filaments driven by myosin V agreed with the helical movement
of single molecules mentioned above, myosin II interestingly showed a
left-handed twirling motion as well.  This result came as a surprise and the
left-handed rotation is opposite from the observations by Nishizaka et
al. \cite{nishizaka93}.  But they are not in direct contradiction, as they
were obtained with quite different ATP concentrations.

While the pitch of the twirling motion is a direct measure for the step size of
processive motors, its interpretation is more complicated with non-processive
ones. They could clearly generate twirling motion if there was a lateral
component of the power stroke. In fact, there exists indirect evidence for such
an asymmetry in myosin V \cite{Purcell.Spudich2005}.  However, we will show in
this paper that there is another, more subtle, effect that can cause twirling
motion of actin filaments in a gliding assay, even if the myosin heads exhibit
no lateral motion.  This effect stems from the fact that myosin heads can only
bind to an actin filament in so called target zones, where the actin binding
sites have approximately the right orientation (Fig.~\ref{fig:1}D)
\cite{Steffen.Sleep2001,Capitanio.Bottinelli2006}.  When a target zone is
approaching a myosin head, the latter is more likely to bind at the beginning
of the target zone than at its end, because it is more likely that it is
already bound by that time. In this paper we will show simulation results and
develop an approximative theory to estimate the pitch of helical motion
resulting from this effect.  Of course, we cannot exclude that there are other
contributions towards the helicity.  But, because the rotation is relatively
weak as compared with the longitudinal motion, these effects can easily be
treated separately and the total rotation is simply their superposition.

\section*{Model definition}

In order to concentrate on the effect of filament rotation, we define a model
with simplified myosin kinetics, essentially containing a detached state, a
bound pre-powerstroke state, a bound post-powerstroke state with ADP and a
bound rigor state. At the same time, we take into account the full helical
structure of the actin filament.  We propose an actin filament which can move
in one direction and rotate around its axis. The position of the actin filament
at a given time is therefore described with the coordinates $(X,\Theta)$. As
follows from the helical structure of the actin filament, the $x$-coordinate of
each bindings site is $X+ia$ and its azimuth angle $\Theta+i \vartheta_0$,
where $\vartheta_0=-(13/28)\times 360^\circ=-167.14^\circ$ is the
rotation and $a=2.75\,\rm nm$ the axial rise per subunit.  The definition of
the model is illustrated in Fig.~\ref{fig:1}.

\begin{figure}
    \includegraphics{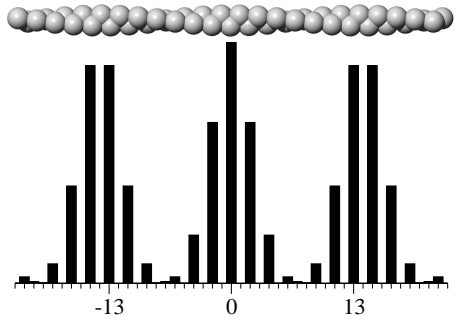}
  \caption{Target zones on an actin filament. The diagram shows the binding
    rate to the particular site if there is no longitudinal strain, i.e., if
    the head is horizontally aligned with the binding site.  The binding rate
    has its maxima for actin subunits oriented downwards (towards the
    myosin-coated surface). The dimensionless angular stiffness is $\alpha=4$.}
\label{fig:2}
\end{figure}

We assume that the myosin motors are distributed randomly directly under the
gliding actin filament (a discussion how this simplified, one-dimensional
model follows from the full, 2-D model is given in the Appendix). The motor
numbered $j$ is anchored at position $x_M^j$. The elastic energy cost of
binding a head to the site $i$ consists of a longitudinal component with
stiffness $K$ and and an angular component with stiffness $K_\vartheta$ and can
be written as
\begin{equation}
\label{eq:1}
  U_i=\frac12 K (X+ia-x^j_M)^2 + \frac12 K_\vartheta (\Theta+i \vartheta_0
  +2 \pi n )^2
\end{equation}
with $n$ chosen such that the angle $\Theta+i\vartheta_0+2\pi n$ falls
into the interval $[-\pi,\pi]$.  
The binding rate is then proportional to the Boltzmann factor
\begin{equation}
  \label{eq:2}
  k_A^i=k_A \exp\left[ \frac{U_i}{k_BT} \right]\;.
\end{equation}
This is essentially the expression used by Steffen et
al. \cite{Steffen.Sleep2001} to fit binding rates of a single myosin
head to the actin filament. They determined the value of the angular
stiffness expressed with the dimensionless coefficient
\begin{equation}
\alpha=K_\vartheta/k_BT
\end{equation}
as $\alpha=3.7$.  However, this value needs to be regarded as a lower
estimate, as it might partially result from torsional compliance of
the actin filament, rather than myosin heads. We therefore use three
different values of $\alpha=4,6$ and $8$ in the simulation. The
angular contribution to the Boltzmann factor for a set of binding
sites is shown in Fig.~\ref{fig:2}. For the longitudinal
compliance, we use the value $K=0.5\,\rm pN/nm$, somewhat below the
stiffness of myosin heads in muscle, which is about $2.5\,\rm pN/nm$
\cite{vilfan2003b}.  The lower stiffness reflects the additional
compliance due to myosin tails and roughly corresponds to the value
obtained with optical tweezers, $0.69\pm 0.47\,\rm pN/nm$
\cite{veigl98}.

The force and the torque that a myosin head numbered $j$, bound to
site $i$, exerts on the filament are
\begin{equation}
  \label{eq:4}
  F^j=K(x^j_M+\delta-X-ia)\qquad M^j=-K_\vartheta(\Theta+i\vartheta_0+2\pi n)\;.
\end{equation}
Here we introduced the displacement $\delta$ that has the value $0$ in
the pre-powerstroke and $d=8\,\rm nm$ in the post-powerstroke and rigor 
state.

The simplified model for the duty cycle of the myosin head is defined as
follows.  A head binds to an actin site with the rate $k_i$ given by
Eq.~(\ref{eq:2}). The power stroke (a transition from $\delta=0$ do
$\delta=d$) takes place with the rate $k_{PS}$.  It is followed by the
release of ADP with the rate $k_{-ADP}$. Detachment follows after binding a
new ATP molecule, therefore its rate depends on the ATP concentration,
$k_D=k_D^0\rm [ATP]$. We neglect the strain dependence of those rates, as well
as the existence of the reverse transitions.  It should be noted that in this
formulation the model is not thermodynamically consistent.  However, as we are
only interested in dynamics at low loads, this does not significantly affect
the results.

We assume that the filament position is quickly equilibrated after
each step, therefore it always fulfills
\begin{equation}
  \label{eq:5}
  F=\sum F^j=0 \qquad M=\sum M^j=0\;.
\end{equation}

\section*{Simulation results}

Given the known structure of the actin helix and the power stroke size of
myosin, our model essentially has two important parameters: the angular
stiffness of myosin heads, $K_\vartheta$, and the ratio between the detachment-
and the attachment rate, $k_D/k_A$.  The latter is a function of the ATP
concentration and is closely related to the duty ratio of motors.  For other
parameters we use the values given in Table~\ref{tab:1}. The power
stroke, connected with the phosphate (Pi) release, is assigned a very
fast rate, $k_{\rm PS}=10000\,{\rm s}^{-1}$, and can be considered as taking
place immediately after binding.  The maximum attachment rate $k_A$, i.e., the
attachment rate to sites that do not require any elastic distortion, can be
estimated as $50\,\rm s^{-1}$. This reflects the estimated average attachment
rate of $30\,\rm s^{-1}$, or a maximum ATP turnover rate of $25\,\rm s^{-1}$ in
muscle \cite{Howard_book}.  The ADP release takes place with the rate
$k_{-ADP}=1000\,\rm s^{-1}$, characteristic for the fast myosin isoform
\cite{Capitanio.Bottinelli2006}.  For the detachment rate, which is determined
by the ATP binding rate, we use $k_D^0=5\,\rm \mu M^{-1} s^{-1}$
\cite{Capitanio.Bottinelli2006}.

\begin{table}
  \begin{tabular}{lll}\hline
    $k_A$ & $50\,{\rm s}^{-1}$ & Attachment rate\\
    $k_{PS}$ & $10000\,{\rm s}^{-1}$ & Power stroke rate\\
    $k_{-ADP}$ & $1000\,{\rm s}^{-1}$ & ADP release rate\\
    $k_D$ & $(5\,\rm \mu M^{-1} s^{-1}) \rm [ATP]$ & Detachment rate\\
    $d$ & $8\,\rm nm$ & Power stroke size\\
    $K$ & $0.5\,\rm pN/nm$ & Myosin stiffness (longitudinal)\\
    $K_\vartheta=\alpha k_BT$\hspace*{-1cm} &  & Myosin stiffness (angular)\\
    $\alpha$ & 4, 6, 8 & Dimensionless angular stiffness\\
    $a$ & $2.75\,\rm nm$ & Distance between actin subunits\\
    $\vartheta_0$ & $- 167.14^\circ$ & Angle between actin subunits\\ 
    $l$ & $5.5\,\rm \mu m$ & Actin filament length\\
    $\rho$ & $20\,\rm \mu m^{-1}$ & 1-d myosin density on surface\\
    $k_BT$ & $4.14\times 10^{-21}\rm J$ & Thermal energy\\
    \hline
  \end{tabular}
  \caption{Simulation parameters}
  \label{tab:1}
\end{table}

The stochastic simulation essentially followed the following algorithm:
\begin{enumerate}
\item Distribute the positions of myosin motors $x_M^j$ randomly along
  the distance covered by the actin filament, with an average linear
  density $\rho$. Set $X=0$ and $\Theta=0$.
\item Determine the total rate of all possible transitions as
  \begin{multline*}
    k_{\rm total}=\sum_{j\in \text{Motors}}\\ \left\{\begin{array}{ll}
        \sum_{i\in \text{Binding sites}} k_A^i(j) & \text{if motor $j$ detached}\\
        k_{PS} &  \text{if motor $j$ in pre-PS state}\\
        k_{-ADP} &  \text{if motor $j$ in post-PS state}\\
        k_{D} & \text{if motor $j$ in rigor state}
\end{array}\right.
\end{multline*}
$k_A^i(j)$ is determined using Eq.~(\ref{eq:2}) with the
current values of $X$ and $\Theta$.
\item Determine the time until the next step as $\Delta t=k_{\rm
    total}^{-1} \ln (1/r)$, where $r$ is a random number between $0$
  and $1$.
\item Choose randomly one of the possible steps (attachment, detachment, power
  stroke, ADP release), so that the probability of choosing a certain
  step is given by its rate, divided by $k_{\rm total}$.
\item Change the state of the chosen motor and update the filament
  position $X$ and angle $\Theta$ according to
  Eq.~(\ref{eq:5}).
\item Continue with step 2 until $X\ge X_{\rm max}$ ($X_{\rm
    max}=1000 \,\rm \mu m$).
\item Determine the average speed as $v=X/t$ and pitch as
  $\lambda=2\pi X/\Theta$.
\end{enumerate}

\begin{figure}
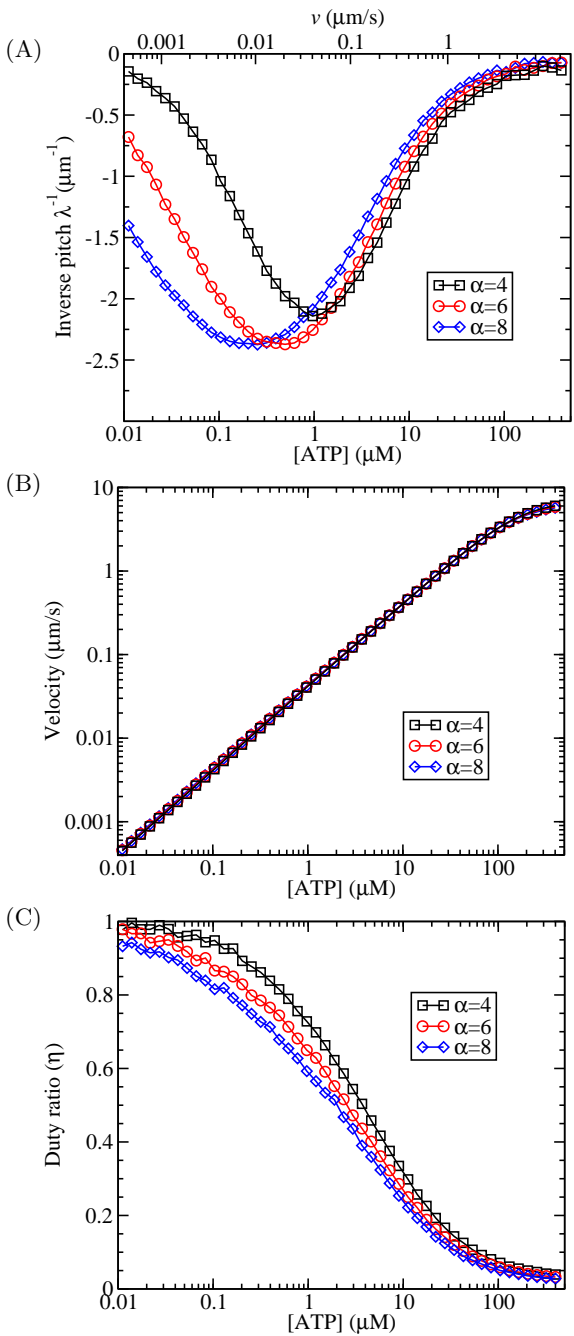

\begin{flushleft}
\raisebox{5.5cm}{(A)}\includegraphics{Figure3a}\\
\raisebox{5.5cm}{(B)}\includegraphics{Figure3b}\\
\raisebox{5.5cm}{(C)}\includegraphics{Figure3c}
\end{flushleft}
\caption{(A) Inverse twirling pitch (number of rotations per micron traveled)
  as a function of the ATP concentration.  The lines represent 3 different
  values of the angular stiffness ($\alpha=4$, $\alpha=6$ and $\alpha=8$).
  Negative signs denote left-handed rotation. The upper scale shows the
  velocity (for $\alpha=4$). (B) Gliding velocity as a function of the
  detachment rate. (C) Duty ratio $\eta$, giving the average fraction of myosin
  heads in the bound state.}
\label{fig:3}
\end{figure}

The results of this numerical simulation are shown in Fig.~\ref{fig:3}, which
shows the inverse pitch of twirling as a function of the ATP concentration for
three different values of the angular stiffness $K_\vartheta$.  For reference,
velocity (Fig.~\ref{fig:3}B) and duty ratio (Fig.~\ref{fig:3}C) are included as
well.  The behavior of the pitch is non-monotonous: it has a minimum of about
$400-500\,\rm nm$ at intermediate speeds, but increases both at high, as well
as very low speeds.

These results show that the helicity of the actin filament is sufficient to
explain the twirling motion in a gliding assay.  Somewhat counter-intuitively,
this rotation is left-handed, and therefore opposite from the handedness of the
actin filament.  The effect becomes weaker for high speeds (where the distance
traveled between two attachment events becomes longer), as well as for very low
speeds, where motors have enough time to bind even to unfavorable sites outside
the target zones.

\begin{figure}
\includegraphics{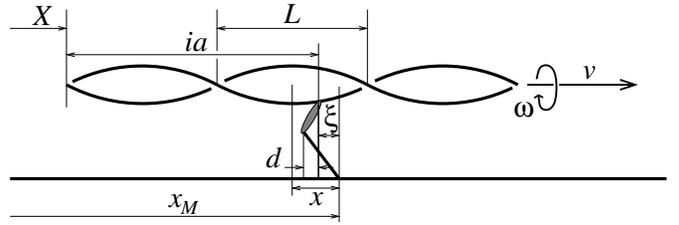}
  \caption{Simplified model used for the analytical solution. The filament
    travels with a stationary velocity $v$ and rotates with an angular velocity
    $\omega$. $x$ denotes the position of a myosin anchoring point relative to
    the center of the target zone.}
  \label{fig:4}
\end{figure}

\section*{Analytical approximation}

In the following we will describe an approximative analytical solution
with the aim of understanding and quantitatively reproducing the
twirling dynamics.  The essential simplification we will make is to
neglect the discrete nature of binding sites on the actin filament and
replace them with a continuous helical ``groove''.  The simplified
model is shown in Fig.~\ref{fig:4}.  We denote each head with
its root position $x$ relative to the center of the target zone:
\begin{equation}
  \label{eq:6}
  x=x_M-\left(X-\frac{L}{\pi}\Theta + nL\right)
\end{equation}
with $n$ such that $-\frac L 2 < x < \frac L 2$.  A bound head is additionally
characterized by the strain $\xi$, which is the position of its root relative
to its binding site.  For for a head bound to site $i$, this is
$\xi_i=x_M-X-ia$.

In the original model the total binding rate for a motor positioned at $x$ is
\begin{equation}
  \label{eq:7}
  \bar k_A(x)=\sum_i k_A^i(x)=\sum_i k_A \exp\left[ -\frac{K\xi_i^2+ K_\vartheta \vartheta_i^2 }{2k_BT} \right] \;.
\end{equation}
In the sum, we only consider sites that are turned towards the motor-covered
surface, therefore we can write the angle as 
\begin{equation}
  \label{eq:8}
\vartheta_i=(x-\xi_i)\frac \pi L
\end{equation}
and by completing the square in the numerator we obtain
\begin{multline}
  \label{eq:9}
k_A^i(x)= k_A(x,\xi_i)\\=k_A \exp\left[
    -\frac{(K+K_\vartheta')(\xi_i-\frac{K_\vartheta'}{K+K_\vartheta'}x)^2+ \frac{K K_\vartheta'}{K+K_\vartheta'} x^2}{2k_BT}
  \right]\;.
\end{multline}
In this equation we introduced the reduced angular stiffness
$K_\vartheta'=( {\pi^2}/{L^2}) K_\vartheta$.  When we neglect the
discreteness of binding sites and extend the summation beyond one
period, we can replace the sum by an integral
\begin{align}
  \label{eq:10}
  \bar k_A(x)&\approx \frac 1 {2a} \int_{-\infty}^{\infty} k_A(x,\xi)d\xi\notag
  \\ &=
\frac {k_A} {2a} \sqrt{\frac{2\pi k_B T}{K+K_\vartheta'}}
  \exp\left[ - \frac{ K K_\vartheta'/(K+K_\vartheta') x^2}{2k_BT} \right]\;.
\end{align}
For a fixed $x$, the the expected value of the strain at the time of
attachment can be calculated using (\ref{eq:9}):
\begin{equation}
  \label{eq:11}
  \left< \xi_A (x) \right> = \frac{\sum_i k_A^i (x)\xi_i}{\sum_i k_A^i(x)}\approx \frac{\int_{-\infty}^\infty k_A (x,\xi)\xi d\xi}{\int_{-\infty}^\infty  k_A(x,\xi)d\xi}  = \frac{K_\vartheta'}{K+K_\vartheta'} x
\end{equation}
The expected value of the azimuthal angle at time of attachment
follows from Eq.~(\ref{eq:8})
\begin{equation}
  \label{eq:12}
   \left< \vartheta_A (x)\right>= \frac{K}{K+K_\vartheta'} \left( \frac{\pi}{L} \right)   x \;.
\end{equation}

\begin{table}
  \begin{tabular}{lp{0.8\columnwidth}}
    \hline
    $L$ & period (half-pitch) of the actin superhelix\\
    $x$ & motor root position relative to the center of the nearest target
    zone\\
    $\xi$ & motor root position relative to its binding site on actin\\
    $k_A^i$ & attachment rate to site $i$\\
    $\bar k_A(x)$ & total attachment rate for a motor positioned at $x$\\
    $\left< \xi_A(x)\right>$ & average strain of newly attached motors
    positioned at  $x$\\
    $\left< \xi_A\right>$ & average strain of all newly attached motors\\
    $\left< x_A\right>$ & average position of newly attached motors,
    relative to the target zone\\
    $\left< \vartheta_A\right>$ & average angular strain at attachment\\
    $v$ & filament velocity\\
    $c$ & apparent velocity of the actin helix\\
    $\omega$ & angular velocity of actin rotation\\
    \hline
  \end{tabular}
  \caption{Variables used in the analytical calculation}
  \label{tab:2}
\end{table}

In the stationary state, the filament moves with velocity $\dot X = v$
and rotates with angular velocity $\dot \Theta = \omega$.  From
Eq.~(\ref{eq:6}) it follows that $\dot x=-(v -
\frac{L}{\pi} \omega)=-c$.  This is the apparent velocity with which the
helix moves along the surface.

\begin{figure}
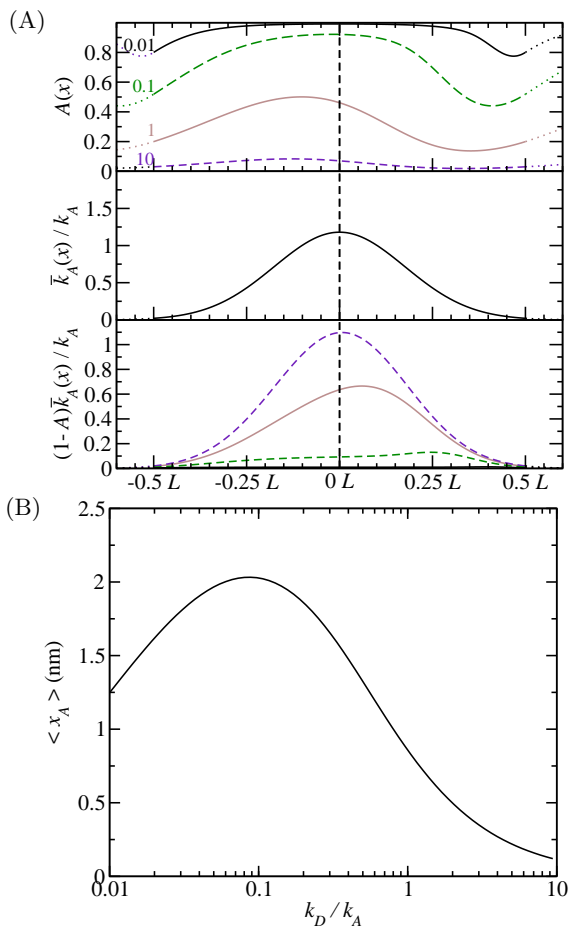

\raisebox{6.2cm}{(A)}\includegraphics{Figure5a}\\
\raisebox{5.5cm}{(B)}\includegraphics{Figure5b}
  \caption{(A) Top panel: probability $A(x)$ that a motor with a position $x$
    relative to the center of a target zone is in the bound state.  The
    parameters are $k_D/k_A=0.01,0.1,1,10$ (each a different line type) and
    $\alpha=4$. Middle panel: dimensionless attachment rate ${\bar
      k}_A(x)/k_A$. Bottom panel: attachment flux, obtained as the product of
    the probability that a motor is in the detached state and the attachment
    rate.  Note that the asymmetry is most pronounced for $k_D/k_A=0.1$. (B)
    The expectation value of $x$ at the time of binding, $\left< x_A \right>$,
    as a function of $k_D/k_A$.}
\label{fig:5}  
\end{figure}

We can now set up a Master equation for the probability that a motor positioned at $x$ is in the attached state
\begin{equation}
  \label{eq:13}
  \partial_t A(x,t)-c \partial_x A(x,t)=\bar k_A(x) (1-A(x,t)) - k_D A(x,t)
\end{equation}
and set $\partial_t A=0$ to obtain the stationary solution. $A$ also
has to fulfill the periodic boundary condition
\begin{equation}
  \label{eq:14}
  A(-L/2)=A(L/2)\;.
\end{equation}
The expectation value of the attachment position can be calculated as 
\begin{equation}
  \label{eq:15}
\left< x_A \right>= \frac{\int_{-L/2}^{L/2}   x  {\bar k}_A(x) (1-A(x))dx } {\int_{-L/2}^{L/2} {\bar k}_A(x) (1-A(x))dx }
\end{equation}
and the average strain at the time of attachment follows from (\ref{eq:11}):
\begin{equation}
  \label{eq:16}
\left< \xi_A \right>=\frac{K_\vartheta'}{K+K_\vartheta'} \left< x_A \right>\;.
\end{equation}
Because the strain on a motor changes with time as $\dot \xi=-v$, the
average strain of all bound motors is $\left< \xi_A \right> -
v/k_D$. The force per motor is then $F= K\left< \xi + d \right>=K(
\left< \xi_A \right> - v/k_D + d$.  As the total force produced by all
motors has to be zero, we obtain an expression for the velocity
\begin{equation}
  \label{eq:17}
  v=k_D(d+ \left< \xi_A \right> )\;.
\end{equation}
The same type of calculation as for the velocity can be made for the
angular velocity.  Motors attach with an average angle
$\left<\vartheta_A\right>$.  As the filament rotates, their angle
changes as $\dot \vartheta = \omega$. The average angle of all motors
is $\left<\vartheta_A\right>+\omega / k_D$ and needs to be zero because
of torque balance, therefore
\begin{equation}
  \label{eq:18}
  \omega=-k_D \left<\vartheta_A\right> = - k_D \frac{K}{K+K_\vartheta'} \left( \frac{\pi}{L} \right)  \left< x_A \right> \;.
\end{equation}
These equations, together with the Master equation (\ref{eq:13}), the periodic
boundary condition (\ref{eq:14}), the expression for $\left< \xi_A \right>$
(\ref{eq:16}) and for $\bar k_A$ (\ref{eq:7}) allow us to numerically determine
the velocity $v$ and the distribution of attached heads $A(x)$ in a
self-consistent manner.  An example of the solution $A(x)$, along with the
attachment rate $\bar k_A(x)$ and attachment flux $\bar k_A(x) (1-A(x))$ is
shown in Fig.~\ref{fig:5}A.  Well visible is the asymmetry in the attachment
flux. The expectation value of $x$ at the time of attachment ($\left<
  x_A\right>$) as a function of the ratio $k_D/k_A$ is shown in
Fig.~\ref{fig:5}B.  It reaches its maximum when $k_D/k_A$ is such that each
motor travels an average path of $\approx L/3$ between two attachment events.

This finally gives us the expression for the twirling pitch
\begin{equation}
  \label{eq:19}
  \lambda = \frac {2 \pi v}{\omega}= -\frac{2 L d \left( 1 + \frac{K_\vartheta'}{K}\right)} {\left< x_A \right>} - 2 L \frac{K_\vartheta'}{K}\;.
\end{equation}

\begin{figure}
  \begin{center}
    \includegraphics{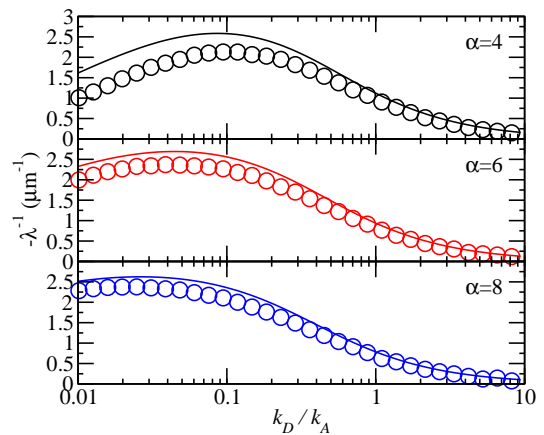}
  \end{center}
  \caption{Inverse pitch $-\lambda^{-1}$, obtained from Eq.~(\ref{eq:19})
    (continuous line), and from simulation described in the previous section,
    however assuming $k_{-ADP}=\infty$. (circles). The top panel shows data for
    the angular stiffness $\alpha=4$, the middle $\alpha=6$ and the bottom
    $\alpha=8$. The minor deviation is mainly due to the extrapolations we made
    in regions between target zones. The agreement is better for large $\alpha$
    values, where target zones become more localized.}
\label{fig:6}  
\end{figure}

The results are shown in Fig.~\ref{fig:6} and compared with
simulation data from Fig.~\ref{fig:3}.  The simulation
results are well reproduced, although there is a certain discrepancy
which is more pronounced for low values of the angular stiffness
$\alpha$.  The main reason for this discrepancy is the extrapolation
beyond the boundaries of one period, which was used in the derivation
of Eq. (\ref{eq:10}).  Other (minor) sources of deviation are
the neglected discrete nature of binding sites and of the fact that
each binding site can only be occupied by one head at a time.

\section*{Discussion}

In this study we have demonstrated that the helical actin structure, along with
the fact that myosin heads preferentially bind to those sites oriented towards
them, is sufficient to explain left-handed rotation in a gliding assay.  The
maximum twirling motion is achieved at relatively low speeds (below $100\,\rm
nm/s$).  Twirling is reduced with higher speeds, achieved at higher ATP
concentrations.  Interestingly, it is also reduced under extremely low ATP
concentrations, when the velocity drops under $10\,\rm nm/s$.  However, in this
regime the results depend strongly on the choice of the angular stiffness
$\alpha$, which is less well known.  The minimum pitch resulting from this
effect lies in the order of $400$--$500\,\rm nm$, which is in good agreement
with recent experimental results \cite{Beausang.Goldman2008}.  The model also
makes a testable prediction that the pitch should increase with a higher ATP
concentration. Because pitch only depends on the ratio between the attachment
and detachment rate, addition of ADP should have the same effect on the pitch
as a reduced ATP concentration that yields the same filament speed.  If this
turns out not to be the case, it will be a strong indication of a lateral
conformational change in the myosin head connected with the release of ADP.

Although the qualitative aspects of our theory are generic and practically
independent of any assumptions other than the helical actin structure, there
are alternative effects that could well contribute to the twirling motion. One
such possibility is that the power-stroke of the myosin head contains a lateral
(azimuthal, off-axis) component.  A similar effect could also result from an
asymmetric attachment rate, which could cause the attached motors to exert a
certain torque on the filament immediately after binding.  Such a torque does
not contradict the laws of thermodynamics because the first bound state is not
in equilibrium with the detached state. A related idea is described by
Beausang et al.\ \cite{Beausang.Goldman2008} as the ``rigor drag model'' in
which heads in the rigor state exert a torque in the opposite direction from
that immediately after binding. This results in a pitch that depends on the
fraction of time spent in the rigor state. The important difference between the
two concepts is that in our model the torque generated by newly attached myosin
heads depends on the ATP concentration, whereas in the rigor drag model this
torque is constant and the variable pitch is caused by different dwell times in
different states.

In any case both the effect described here and the explicit lateral component
of the power stroke will be superimposed.  So it is theoretically possible,
even though the proposition is purely speculative at the moment, that the power
stroke might have the opposite helicity, i.e., it would lead to a right-handed
filament rotation.  In such a case, there could be a cross-over form right
handed motion under high ATP concentrations, to left-handed under low.  This is
could be one possibility to reconcile the recent results
\cite{Beausang.Goldman2008} with those by Nishizaka et al.\ \cite{nishizaka93}.

Recent experiments also revealed rotation of microtubules moved by monomeric
kinesin-1 \cite{Yajima.Cross2005} and Eg5 \cite{Yajima.Nishizaka2008}.  The
theory we presented in this paper is not applicable to microtubules, because
they have no distinct target zones.  Any rotation resulting from an effect of
the kind we describe here would be negligible.  Therefore, as suggested in
\cite{Yajima.Cross2005}, the rotation caused by kinesins has to result from a
lateral (off-axis) component of a power stroke, or from an asymmetry in the
binding rate.

\section*{Appendix}

In the following we will discuss how the simplified model, which assumes that
motors are distributed one-dimensionally underneath the actin filament, and
which we use throughout the main text, relates to the full two-dimensional
model. In the 2-D model, which describes the actual situation in a gliding
assay, myosin heads are distributed all over the glass surface and their
positions are described with two coordinates, $(x_M^j, y_M^j)$. The position of
the actin filament is described with the coordinates $(X,Y)$ and the angle
$\Theta$.  We assume that the filament keeps its direction parallel to the
$X$-axis.  The elastic distortion of the head $j$ binding to the site $i$ can
then be written as
\begin{equation}
  \label{eq:20}
  U_i=\frac12 K(X+ia-x_M^j)^2 + U_A(\Theta +i \vartheta_0, Y-y_M^j)
\end{equation}
where $U_A$ is an unknown function of the azimuth angle of binding site $i$ and
of the lateral position of the filament relative to the motor.

The total binding rate to site $i$ of all motors located at longitudinal
position $x_M$ is then
\begin{multline}
  \label{eq:21}
  k_A(x_M)\\ =\int_{-\infty}^\infty k_A  \exp\left[-\frac{U_i(X+ia-x_M, \Theta +i
      \vartheta_0, Y-y_M)}{k_BT}\right] \\ \times \rho D(x_M, y_M) dy_M
\end{multline}
where $\rho$ is the 2-D surface density of myosin motors and $D(x_M,y_M)$ the
probability that a motor at that position is in the detached state.  If we
assume that $D(x_M,y_M)=D(x_M,2 Y - y_M)$, i.e., that the distribution of
unbound heads is symmetric with respect to the filament, the resulting function
$k_A(x_M)$ has to be symmetric in $\Theta+i\vartheta_0$. We can therefore
approximate it with the expression used in
Eqs.~(\ref{eq:1},\ref{eq:2}).

The average torque generated by a head that binds to site $i$ can be calculated
as
\begin{align}
  \label{eq:22}
  \left< M \right> &=- \frac{ \int_{-\infty}^\infty \exp\left(-U_i/k_BT\right)
    D(x_M, y_M) \left(\partial U_i/\partial \Theta\right) dy_M}{ \int_{-\infty}^\infty\exp\left(-U_i/k_BT\right)
    D(x_M, y_M) dy_M}\notag \\ &= \frac{k_BT}{k_A(x_M)} \frac{d k_A(x_M)}{d \Theta}\;.
\end{align}
This second expression is equivalent to that in the 1-D model. 

The average force generated by a head that binds to site $i$ is determined the
same way:
\begin{equation}
  \label{eq:23}
  \left< F_L \right> =- \frac{ \int_{-\infty}^\infty \exp\left(-U_i/k_BT\right)
    D(x_M, y_M) \left(\partial U_i/\partial Y \right) dy_M}{ \int_{-\infty}^\infty\exp\left(-U_i/k_BT\right)
    D(x_M, y_M) dy_M} \;.
\end{equation}
If $D(x_M,y_M)$ is independent of $y_M$, or, more generally, if it has a
dependence that can be written as a function of $U_i$, the integral in the
numerator is 0 and there is no lateral force.  However, with different
distributions $D(x_M,y_M)$, a small lateral force is possible, so that the
filament could show some sideways motion in the 2-D model.  While the torque
results from a $D(x_M,y_M)$ which is asymmetric in $x_M$, a lateral force needs
asymmetry in both coordinates and is therefore a higher-order effect.

\section*{Acknowledgment}

I thank John Beausang and Yale E. Goldman for stimulating discussions and
helpful comments on the manuscript. This work was supported by the Slovenian
Research Agency (Grant P1-0099).

%%%%%%%%%%%%%%%%%%%%%%%%%%%%%%%%%%%%%%%%%%%%%%%

\end{document}